\documentclass[a4paper,leqno]{CAIMclass}
\newcommand{\DOI}{XXX} % replace XXX with the last three digits of DOI when assigned by editors
\newcommand{\caimeditor}{Associated editor}
\newcommand{\caimrec}{MM DD}
\newcommand{\caimrecf}{MM DD}
\usepackage{amsmath,amssymb}
\usepackage{graphicx} % support the \includegraphics command and options

\begin{document}
%*********************************
\title{Topological Features of Online Social Networks}
\author{{Emilio Ferrara$^1$, Giacomo Fiumara$^2$}}
\address{
$^1$Department of Mathematics, University of Messina, Italy\\
eferrara@unime.it}
\address{$^2$ Department of Physics, Informatics Section, University of Messina, Italy\\
gfiumara@unime.it}

%\begin{communicated}
%\comby{Associated editor}
%\end{communicated}

\begin{abstract}
The importance of modeling and analyzing Social Networks is a consequence of the success of Online Social Networks during last years.
Several models of networks have been proposed, reflecting the different characteristics of Social Networks.
Some of them fit better to model specific phenomena, such as the growth and the evolution of the Social Networks; others are more appropriate to capture the topological characteristics of the networks.
Because these networks show unique and different properties and features, in this work we describe and exploit several models in order to capture the structure of popular Online Social Networks, such as Arxiv, Facebook, Wikipedia and YouTube.
Our experimentation aims at verifying the structural characteristics of these networks, in order to understand what model better depicts their structure, and to analyze the inner \emph{community structure}, to illustrate how members of these Online Social Networks interact and group together into smaller communities.
\end{abstract}


\AMScode{91D30; 05C82; 68R10; 90B10; 90C35}

%**********************************************
\section{Introduction} \label{sec:introduction}
From a scientific standpoint, the emergence of the phenomenon of Online Social Networks (OSNs), which is obtaining an unpredicted success during last years, contributed to keep high the level of attention of the research community on Social Networks.

The large dimension of these networks is a key aspect.
In fact, they represent an incredible source of information on a large-scale, thus allowing studies which were impossible before by adopting classic sociological investigation techniques.
Studying large-scale Online Social Networks, and their evolution, can be useful to investigate similarities and differences with real-life Social Networks.
Moreover, this could help to empirically confirm, or confute, sociological theories such as the ``Small World'' \cite{Travers1969}, the ``Dunbar number'' \cite{dunbar1998grooming}, and so on.
Another advantage of studying OSNs is that connections and hierarchies among users are clearly defined, because they reflect constraints imposed by the platforms.
This simplifies the process of induction of a graph of relations which reflects the structure of the Social Network.
Moreover, it is possible to evaluate structural properties of large-scale networks adopting optimized algorithms and exploiting computational resources, in the light of the paradigm called Social Network Analysis (SNA), a novel branch of Computational Social Sciences.

In this work we face the problem of analyzing the topological features of some popular Online Social Networks, focusing on the characteristics of the graphs which represent these networks.
To do so, we adopt some specific measures, such as the diameter and the degree distribution.
Moreover, we investigate the emerging \emph{community structure} inside these networks.

This paper is organized as follows: Section \ref{sec:related} focuses on related work, presenting classic literature on Social Networks, their analysis -- in particular regarding Online Social Networks -- and the latest works, which define directions of SNA.
Section \ref{sec:features} introduces the key features reflected by Online Social Networks.
In Section \ref{sec:models} we describe the generative models proposed to represent Social Networks, putting into evidence those aspects that could fit well to represent Online Social Networks and those in which they could fail.
Results of our experimentation, presented in Section \ref{sec:results}, depict the topological features of analyzed Online Social Networks.
Section \ref{sec:conclusions} concludes, suggesting some directions for future works.

%%%%%%%%%%%%%%%%%%%%%%%%%%%%%%%%%%%%%%%%%%
\section{Related Work} \label{sec:related}
\subsection{Background in Social Networks and Models}
Literature about Social Network models is rooted in social sciences: in the sixties, Milgram and Travers \cite{Travers1969} analyzed characteristics of real-life Social Networks, conducting several social experiments, and in conclusion, proposing the well known ``Small World'' model (see Section \ref{sub:small-world}).

Another important concept, introduced by Zachary \cite{Zachary1980}, is the \emph{community structure} (see Section \ref{sub:community-structure}).
He analyzed a small real-life social community (i.e., the components of a karate club), defining a model which describes the clusterization of Social Networks via cuts and fissions in sub-groups.

One of the first models has been provided by Erd\H{o}s and R\'{e}nyi \cite{erdos1959random} (see Section \ref{sub:erdos-renyi}), and employs \emph{random graphs} in order to reproduce real networks.
Watts and Strogatz \cite{watts1998collective} furnished a one-parameter model that interpolates between an ordered finite dimensional lattice and a random graph (see Section \ref{sub:watts-strogatz}).
This because they empirically found that real-world Social Networks are well connected and have a short average path length like random graphs, but they also have exceptionally large clustering coefficients, a feature which is not reflected by random graph models.
Barab{\'a}si and Albert \cite{albert1999diameter,albert2002statistical} introduced different models that can be applied to friendship networks, the World Wide Web, business and commerce networks, etc., proving that they all share similar properties (see Section \ref{sub:barabasi-albert}).

%*********************************************
\subsection{Recent Studies and Current Trends}
Some of the current trends in the analysis of Social Networks include:

a) The investigation of the topological features of Social Networks by means of measurements, studying link symmetries, degree distributions, clustering phenomena, groups formations, etc., usually on a large-scale, by analyzing Online Social Networks \cite{Ahn2007,mislove2007measurement}.

b) The study of evolutionary aspects of Social Networks. 
In this context, Kumar et al. \cite{kumar2010structure} defined a generative model which describes the structure of OSNs and their dynamics over the time. 
This model has been compared against actual data, in order to validate its reliability. 
Similarly, Leskovec \cite{leskovec2005graphs}, analyzed evolutionary aspects of Social Networks, trying to describe dynamic and structural features which influence the growth of communities, in particular when considering large Social Networks.

c) The definition of novel graph mining techniques, to face the computational complexity of studying large-scale social graphs with millions nodes and edges. Some authors \cite{Leskovec2006,Gjoka2010} faced the problem of sampling from large graphs adopting different techniques, in order to establish if it is possible to avoid bias of data studying sub-graphs of Social Networks. 
They found that \emph{Random Walks} and \emph{Metropolis-Hastings} algorithms perform better, respectively, for static and dynamic graphs, concluding that samples of size of 15\% of a social graph preserve the most of the properties.

d) The identification of those characteristics of the network that could suggest what nodes are more likely to be connected by trusted relationships (the \emph{link prediction} problem) \cite{Golbeck2006,Liben-nowell2007}. 
This is of a great interest for different commercial reasons.

\subsubsection{Applications of Social Network Analysis research}
Possible applications of information acquired from Social Networks have been investigated by Staab et al. \cite{Staab2005}: methodologies for exploiting discovered data were defined, for marketing purposes, recommendation and trust analysis, etc. 
Recently, several marketing and commercial studies have been applied to OSNs, in particular to discover efficient channels to distribute information \cite{kempe2003maximizing} and users who share similar tastes and preferences in order to suggest them useful recommendations \cite{demeo2011recommendation}.
Our study provides useful information in all these directions, identifying interesting characteristics of Online Social Networks, considering the topological features that could affect how much efficiently nodes and edges could carry information through the networks.

%*************************************
\subsection{Motivations of our Study}
Several motivations to study features of Online Social Networks hold.
On the one hand, results reported in our study have immediate implications, for example in commercial or marketing context, in the engineering of networks, etc.
On the other hand, our achievements enrich the actual panorama of experimental verification of models and theories proposed by Social Sciences about the features and the dynamics of social networks.

%%%%%%%%%%%%%%%%%*****************************************
\section{Features of Social Networks} \label{sec:features}
In this section we put into evidence three key features that characterize Social Networks, i.e., i) the ``Small World'' effect, ii) \emph{scale-free degree distributions} and, iii) emergence of a \emph{community structure}.
During our experimentation, we take into account these features in order to establish if Online Social Networks show these well-known characteristics.

A Social Network can be defined by means of a graph $G=(V, E)$ whose set of vertices $V$ represents nodes of the network (i.e., the individuals), and whose set of edges $E$ represents connections (i.e., the social ties) among nodes of the network.

Social Networks can be modeled by means of weighted/unweighted and directed/undirected graphs, depending on their characteristics.
In the following, we will conveniently adopt those structures that best fit to solve the given problems.
Several features that we are going to consider are related to some background key concepts regarding graph theory, such as shortest paths or diameter of a graph, that are considered as well-known and not introduced in the following for reason of briefness.

%%%%%%%%%%%%%%%%%%%%%%%%%%%%%%%%%%%%%%%%%%%%%%%%%%%%%%%%
\subsection{The ``Small World'' Effect} \label{sub:small-world}
The study of the ``Small World'' effect on Social Networks is rooted in Social Sciences \cite{Travers1969}.
Authors put into evidence that, despite their big dimension, Social Networks usually show a common feature: there exists a relatively short path connecting any pair of nodes within the network.

In fact, a ``Small World'' network is represented by a graph in which most nodes are not reciprocal neighbors each other, but could be reached from each other node by a small number of hops.
The diameter $\ell$, that reflects the ``Small World'' property, scales proportionally to the logarithm of the dimension of the network, which is formalized as 
\begin{equation}
	\ell \propto Log(|V|)
	\label{eq:small-world-property}
\end{equation}

where $|V|$ means the cardinality of $V$.
Some characteristics of many real-world networks are well-modeled by means of ``Small World'' networks, such as OSNs \cite{watts1998collective}, Internet \cite{pastor2001dynamical}, World Wide Web \cite{albert1999diameter}, biological networks \cite{girvan2002community}.

%%%%%%%%%%%%%%%%%%%%%%%%%%%%%%%%%%%%%%%%%%%%%%%
\subsection{Scale-free Degree Distributions}
An important feature that is reflected by several generative models of Social Networks is the degree distribution of nodes. 
This feature characterizes the way the nodes are interconnected each other in the Social Network.
On the one hand, in a random graph (see further) the node degree (i.e., the number of edges the given node is an endpoint of) is characterized by a distribution function $P(k)$ which defines the probability that a randomly chosen node has exactly $k$ edges.
Because the distribution of edges in a random graph is aleatory, the most of the nodes have approximatively the same node degree, similar to the mean degree $\langle k \rangle$ of the network.
Thus, the degree distribution of a random graph is well described by a Poisson distribution law, with a peak in $P(\langle k \rangle)$.
On the other hand, recent empirical results show that in the most of real-world networks the degree distribution significantly differs from a Poisson distribution.
In particular, for several large-scale networks, such as the World Wide Web \cite{albert1999diameter}, Internet \cite{faloutsos1999power}, metabolic networks \cite{jeong2000large}, etc., the degree distribution follows a power law

\begin{equation}
	\label{eq:powerlaw}
	P(k)\sim k^{-\gamma}
\end{equation}

This power law distribution sensibly differs from Poisson distributions, but, on the other hand, falls off more gradually than an exponential one, allowing for a few nodes of very large degree to exist.
Power law based models (see Section \ref{sub:barabasi-albert}) apparently well depict the node degree distributions of large-scale Social Networks.
Since these power laws are free of any characteristic scale, such a network with a power law degree distribution is called a \emph{scale-free network} \cite{barabasi1999emergence}.

%******************************************************************************
\subsection{Emergence of a Community Structure} \label{sub:community-structure}
Another aspect to take into account when studying Social Networks is the emergence of a \emph{community structure}: the more this structural characteristic is evident, the higher a network tends to divide into groups of nodes whose connections are denser among entities belonging to the given group and sparser otherwise.
Not all the network models are able to represent this characteristic.
For example, the Erd\H{o}s-R\'{e}nyi (see Section \ref{sub:erdos-renyi}) or the Barab\'{a}si-Albert models (see Section \ref{sub:barabasi-albert}) can not meaningfully represent the concept of \emph{community structure}, that emerges from the empirical analysis of Social Networks.
The \emph{community structure} in Online Social Networks is discussed in Section \ref{sec:community-structure}.

%*****************************************************
\section{Models of Social Networks} \label{sec:models}
In literature, different models have been presented based on the previously discussed features.
In this work we focus on the three most widely exploited modeling paradigms: i) random graphs, ii) ``Small World'' networks and, iii) power law networks.
Random graphs represent an evolution of the Erd\H{o}s-R\'{e}nyi model, and are widely used in several empirical studies, because of their ease of adoption.
After the discovery of the ``Small World'' effect, a new class of models, namely ``Small World'' networks, has been introduced.
Similarly, the power law degree distribution emerging from real-world Social Networks led to the modeling of the homonym networks, which are adopted to describe scale-free behaviors and other non-Poisson degree distributions.

\begin{figure}%
	\centering
	\tiny
	a)\includegraphics[width=.45\columnwidth]{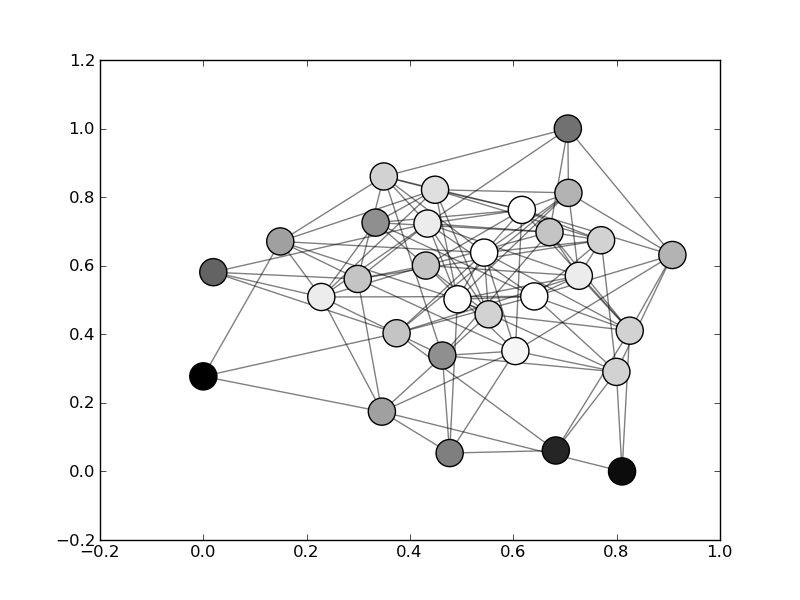}%
	b)\includegraphics[width=.45\columnwidth]{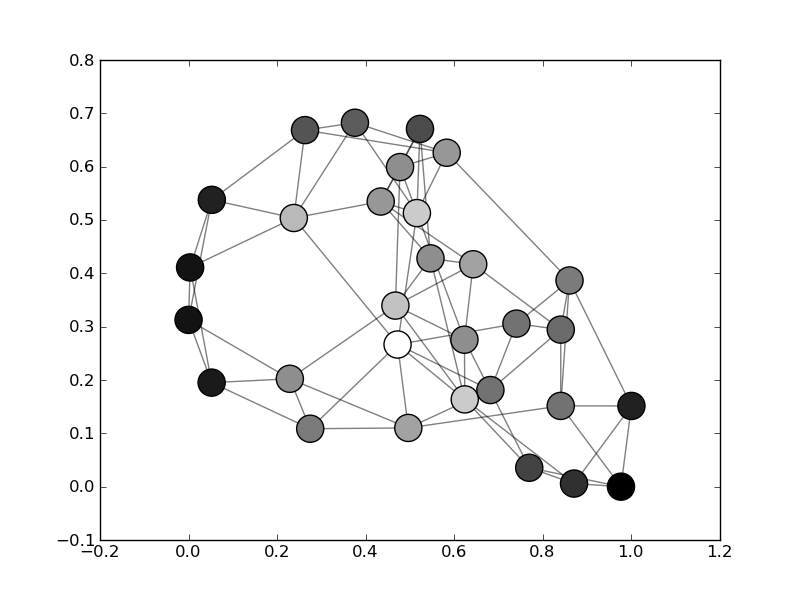}%
	
	c)\includegraphics[width=.3\columnwidth]{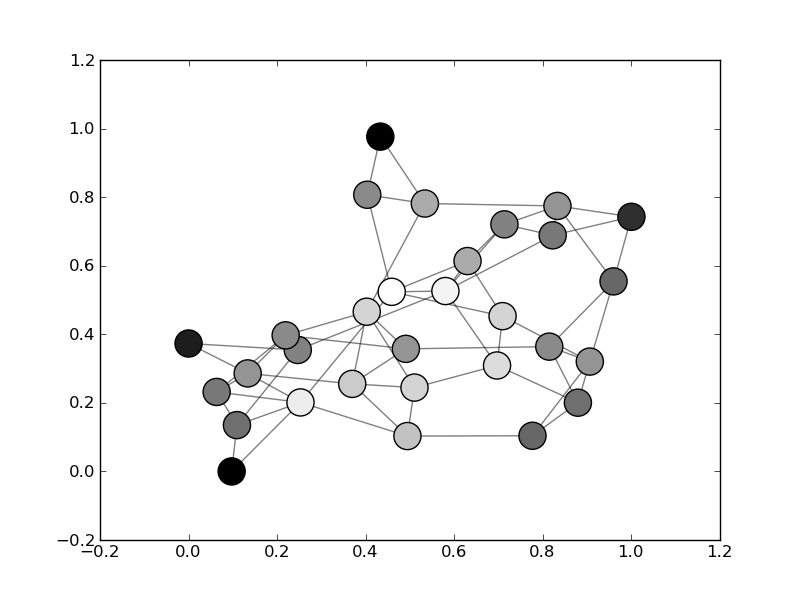}%
	d)\includegraphics[width=.3\columnwidth]{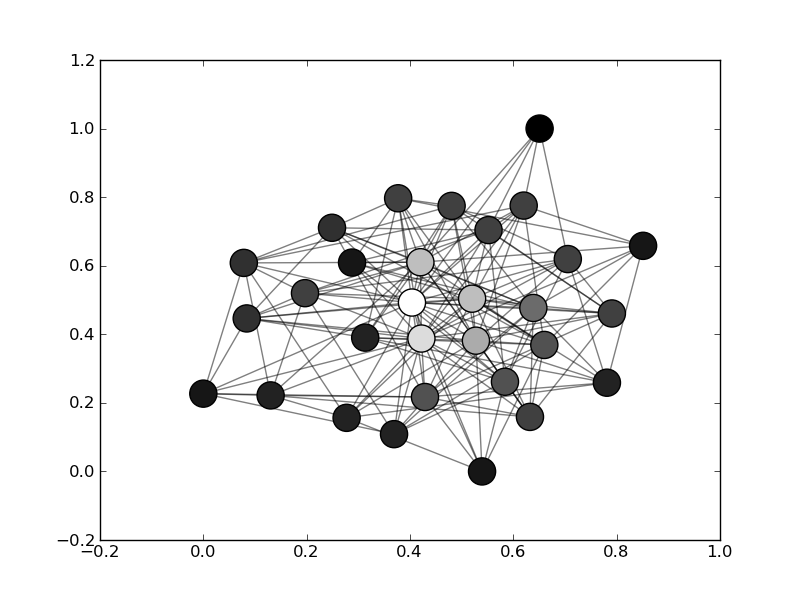}%
	e)\includegraphics[width=.3\columnwidth]{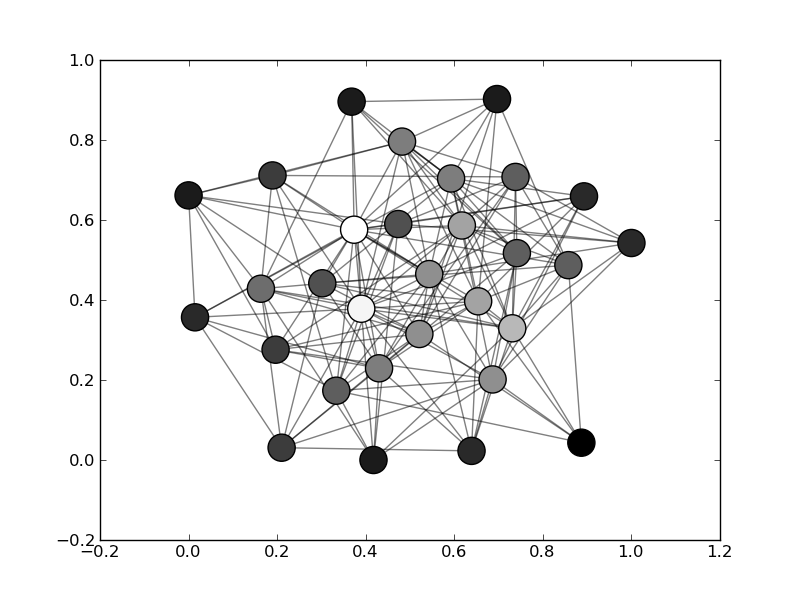}%
	\caption{Generative models of Social Networks: a) Erd\H{o}s-R\'{e}nyi \cite{erdos1959random}; b) Newman-Watts-Strogatz \cite{newman1999renormalization}; c) Watts-Strogatz \cite{watts1998collective}; d) Barab\'{a}si-Albert \cite{barabasi1999emergence}; e) Holme-Kim \cite{holme2002growing}.
	 Lighter nodes have higher closeness centrality \cite{sabidussi1966centrality}, darker nodes have lower closeness centrality. 
	 Nodes are placed according to a spring layout (Fruchterman-Reingold algorithm \cite{fruchterman1991graph}). 
	}%
	\label{fig:models}%
\end{figure}

%*****************************************************************
\subsection{The Erd\H{o}s-R\'{e}nyi Model} \label{sub:erdos-renyi}
Erd\H{o}s and R\'{e}nyi \cite{erdos1959random} proposed one of the first modeling paradigm for networks, the random graph.
They defined two models: the simple one consists of a graph containing $n$ vertices connected randomly.
The commonly adopted model, indeed, is defined as a graph $G_{n,p}$ in which each possible edge between two vertices may be included in the graph with the probability $p$ (and may not be included with the probability ($1-p$)).

Although random graphs have been widely adopted because their properties ease the work of modeling networks (for example, random graphs have small diameters), they do not properly reflect the structure of real-world large-scale networks, mainly for two reasons: i) the degree distribution of random graphs follows a Poisson law, which substantially differs from the power law distribution shown by empirical data; ii) they do not reflect the clustering phenomenon, considering all the nodes of the network with the same \emph{weight}, and reducing, de facto, the network to a giant cluster.

This emerges by considering Figure \ref{fig:models}.a, where is shown a Erd\H{o}s-R\'{e}nyi graph generated by adopting $n=30$ and $p=0.25$. 
The most of the nodes have similar closeness centrality (that is positively correlated to the degree), identified by the gray color in a gray-scale, and this means that all the nodes have relatively similar features (which is consistent with the formulation of the graph model, according with the Poisson distribution of node degrees).
Social Networks exhibit a rather different behavior, making this model unfeasible for modern studies although it has been widely adopted in the past.

%***************************************************************
\subsection{The Watts-Strogatz Model} \label{sub:watts-strogatz}
The real-world Social Networks are well connected and have a short average path length like random graphs, but they also have exceptionally large clustering coefficients, a characteristic that is not reflected by the Erd\H{o}s-R\'{e}nyi model or by other random graph models.
Watts and Strogatz \cite{watts1998collective} proposed a one-parameter model that interpolates between an ordered finite dimensional lattice and a random graph.
Starting from a ring lattice with $n$ vertices and $k$ edges per vertex, each edge is rewired at random with probability $p$, ranging from $0$ (regular network) to $1$ (random network).
Focusing on two quantities, namely the characteristic path length $L(p)$ (defined as the number of edges in the shortest path between two vertices) and the clustering coefficient $C(p)$, some authors \cite{holland1971transitivity} found that $L \sim n/2k \geq 1$ and $C \sim 3/4$ as $p\rightarrow 0$, while $L=L_{random} ln(n)/ln(k)$ and $C=C_{random} k/n \leq 1$ as $p \rightarrow 1$.
The Watts-Strogatz model is therefore suitable for explaining such properties in many real-world examples.

This model has been widely studied since the details have been published.
The disadvantage of the model, however, is that it is not able to capture the power law degree distribution as presented in most real-world Social Networks.

A strong structural difference is evident between Watts-Strogatz \cite{watts1998collective}, its variant Newman-Watts-Strogatz \cite{newman1999renormalization} presented in Figures \ref{fig:models}.b and \ref{fig:models}.c if compared with the Erd\H{o}s-R\'{e}nyi graph (Figure \ref{fig:models}.a).
First of all, it emerges that centrality of nodes is more heterogeneous, covering the whole gray-scale. 
On the other hand, is evident, if compared with the other models, that it can not well reflect the power law distribution of node degree experimentally shown by data, even if a \emph{community structure} is well represented (see Section \ref{sec:community-structure}).

%*********************************************************************
\subsection{The Barab\'{a}si-Albert Model} \label{sub:barabasi-albert}
The two previously discussed theories observe properties of real-world networks and attempt to create models that incorporate those characteristics.
However, they do not help in understanding the origin of Social Networks and how those properties evolve.

The Barab\'{a}si-Albert model \cite{barabasi1999emergence,albert2002statistical} suggests that two main ingredients of self-organization of a network in a scale-free structure are \emph{growth} and \emph{preferential attachment}.
These pinpoint to the facts that the most of networks continuously grow by the addition of new nodes which are preferentially attached to existing nodes with large numbers of connections.
The generation scheme of a Barab\'{a}si-Albert scale-free model is as follows: (i) \emph{Growth}: let $p_k$ to be the fraction of nodes in the undirected network of size $n$ with degree $k$, so that $\sum_k p_k = 1$ and therefore the mean degree $m$ of the network is $\frac{1}{2} \sum_k kp_k$.
Starting with a small number of nodes, at each time step, we add a new node with $m$ edges linked to nodes already part of the system;
(ii) \emph{preferential attachment}: the probability $\prod_i$ that a new node will be connected to the node $i$ (one of the $m$ already existing nodes) depends on the degree $k_i$ of the node $i$, in such a way that $\prod_i=k_i \sum_j k_j$.

Models based on preferential attachment operates in the following way.
Nodes are added one at a time.
When a new node $u$ has to be added to the network it creates $m$ edges ($m$ is a parameter and it is constant for all nodes).
The edges are not placed uniformly at random but {\em preferentially}, i.e., the probability that a new edge of $u$ is placed to a node $v$ of degree $d(v)$ is proportional to its degree, $p_u(v) \propto d(v)$.
This simple behavior leads to power law degree tails with exponent $\gamma = 3$.
Moreover it also leads to low diameters.
While the model captures the power law tail of the degree distribution, it has other properties that may or may not agree with empirical results in real networks.
Recent analytical research on average path length indicate that $\ell \sim ln(|V|)/lnln(|V|)$.
Thus the model has much shorter $\ell$ with respect to a random graph.
The clustering coefficient decreases with the network size, following approximately a power law $C \sim N^{-0.75}$.
Though greater than those of random graphs, it depends on the size of the network, which is not true for real-world Social Networks.

Figures \ref{fig:models}.d and \ref{fig:models}.e propose two example of graphs generated by using the Barab\'{a}si-Albert scale-free model \cite{barabasi1999emergence} and a variant by Holme and Kim \cite{holme2002growing}.
It is evident that the structure of these networks is much more compact than the Watts-Strogatz models (Figures \ref{fig:models}.b and \ref{fig:models}.c) but there are a few nodes that have a very high centrality while the most of the others have very low degrees (those depicted in dark gray).
Due to the spring layout given by the Fruchterman-Reingold algorithm \cite{fruchterman1991graph}, those nodes with low degrees (belonging to the tail of the power law) are represented in peripheral positions in respect to central nodes.
On the other hand, this model fails in representing a meaningful \emph{community structure} of the network, differently to the Watts-Strogatz based models (see further).

%*************************************************************
\subsection{Discovering Communities} \label{sub:models-communities}
Several studies have been conducted in order to investigate the \emph{community structure} of real and Online Social Networks \cite{karrer2008robustness,porter2009communities,fortunato2010community}.

In its general formulation, the problem of finding communities in a network is intended as a clustering problem, thus solvable by assigning each vertex of the network to a cluster, in a meaningful way.
There are essentially two different and widely adopted approaches to solve this problem; the first is the spectral clustering \cite{ng2001spectral} which relies on optimizing the process of cutting the graph; the latter is based on the concept of \emph{network modularity}.

The problem of minimizing the graph-cuts is NP-hard, thus an approximation of the exact solution can be obtained using the spectral clustering \cite{ng2001spectral}, exploiting the eigenvectors of the Laplacian matrix of the network.
This process can be performed using the concept of ratio cut \cite{hagen2002new}, a function which can be minimized in order to obtain large clusters with a minimum number of outgoing interconnections among them. 
The main limitation of the spectral clustering is that the number of communities present in the network and their size need to be defined in advance. 
This makes it unsuitable if the purpose is to discover the underlying community structure of an unknown network.

The network modularity concept can be explained as follows: let consider a network, represented by means of a graph $G=(V,E)$, which has been partitioned into $m$ communities; its value of network modularity is 
\begin{equation}
	\label{eq:qmod}
	Q= \sum_{s = 1}^m \left[\frac{l_s}{|E|} - \left(\frac{d_s}{2|E|}\right)^2\right]
\end{equation} 
assuming $l_s$ the number of edges between vertices belonging to the $s$-th community and $d_s$ the sum of the degrees of the vertices in the $s$-th community.
Intuitively, high values of $Q$ implies high values of $l_s$ for each discovered community; thus, detected communities are dense within their structure and weakly coupled among each other. 
Because the task of maximizing the function $Q$ is NP-hard, several approximate techniques have been presented during the last years.

Let us consider the Girvan-Newman algorithm \cite{girvan2002community,newman2004finding,newman2006modularity}.
It first calculates the {\em edge betweenness} $B(e)$ of a given edge $e$ in $G=(V,E)$, defined as
\begin{equation}	
	B(e) = \sum_{n_i \in V}\sum_{n_l \in V}\frac{np_e(n_i,n_l)}{np(n_i,n_l)}
	\label{eq:bc}
\end{equation}
where $n_i$ and $n_l$ are vertices in $V$, $np(n_i,n_l)$ is the number of the shortest paths between $n_i$ and $n_l$ and $np_e(n_i, n_l)$ is the number of the shortest paths between $n_i$ and $n_l$ containing $e$.
It is possible to maximize the value of $Q$ deleting edges with a high value of betweenness, because connecting vertices belonging to different communities.
Starting from this intuition, first the algorithm ranks all the edges with respect to their betweenness, thus removing the most influential, calculates the value of $Q$ and iterates the process until a significant increase of $Q$ is obtained.
At each iteration, each connected component of $V$ identifies a community.
Its cost is $O(n^3)$, being $n$ the number of vertices in $V$; intuitively, it is unsuitable for large-scale networks.

Several variants and improvements exist; in this work, in order to discover the \emph{community structure} of analyzed Online Social Networks, we the \emph{Louvain method} \cite{blondel2008fast}.
At the best of our knowledge, does not exist a model of Social Network based on the modularity optimization, and this aspect leaves space for further investigations.

\begin{figure}%
	\centering \tiny
	a)\includegraphics[width=.45\columnwidth]{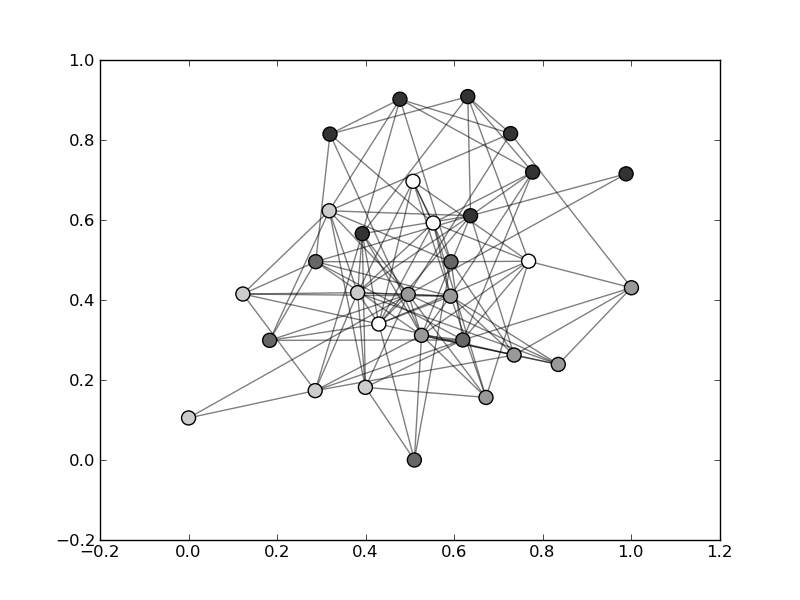}%
	b)\includegraphics[width=.45\columnwidth]{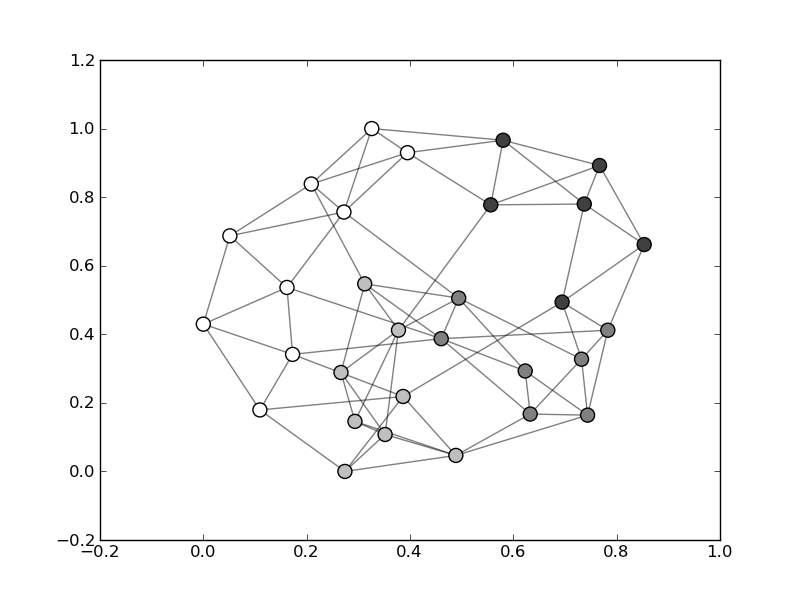}%
	
	c)\includegraphics[width=.3\columnwidth]{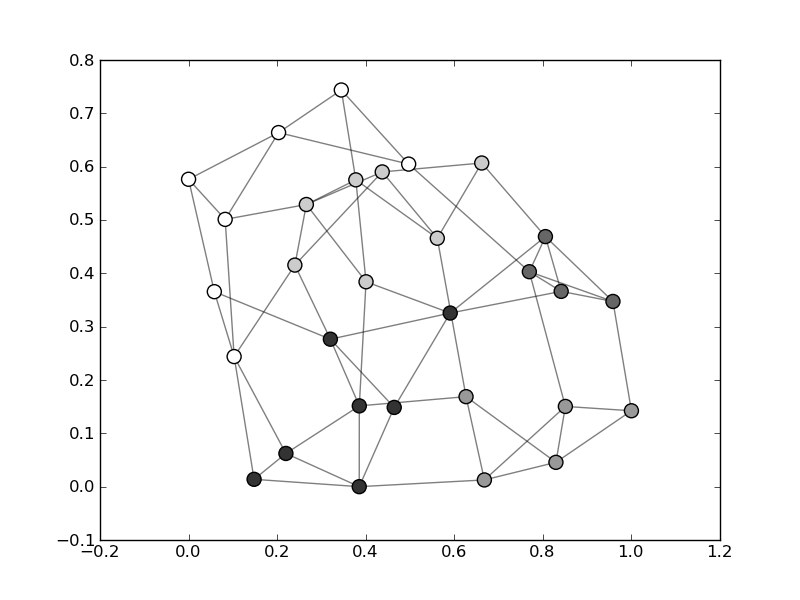}%
	d)\includegraphics[width=.3\columnwidth]{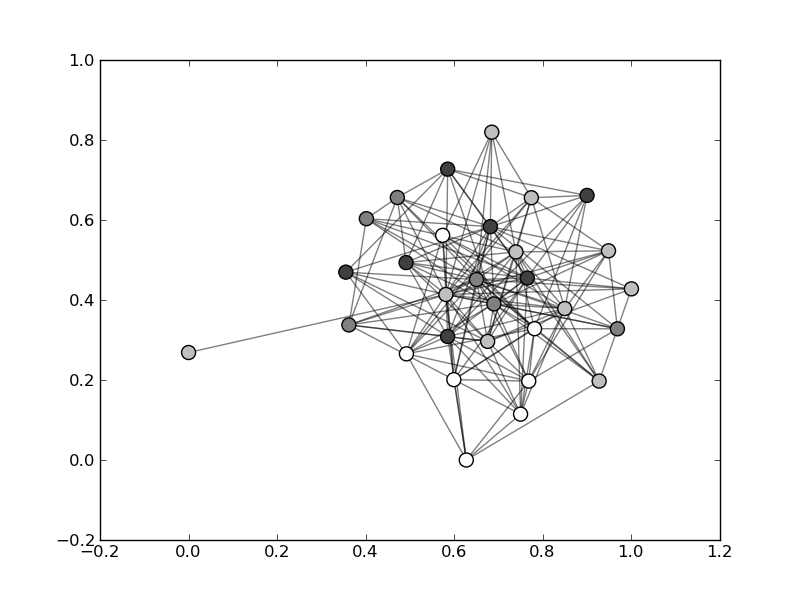}%
	e)\includegraphics[width=.3\columnwidth]{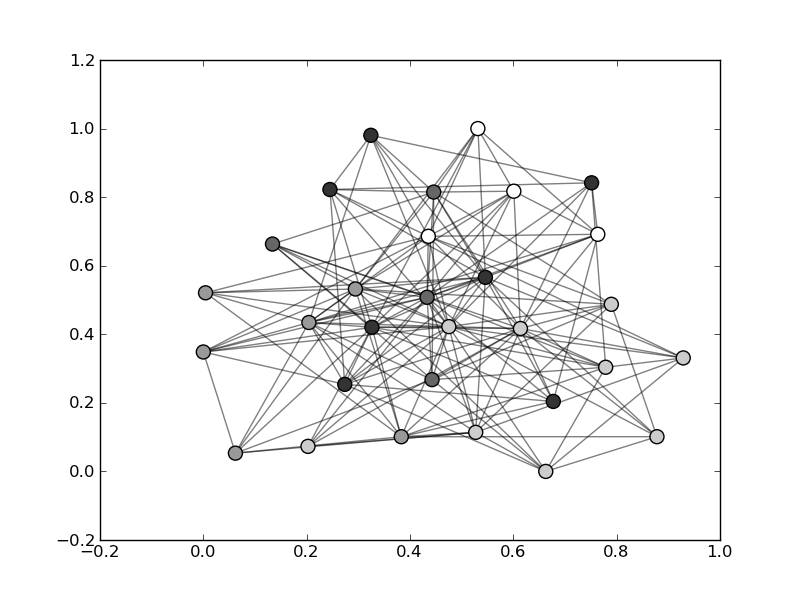}%
	\caption{\emph{Community structure} of models of Social Networks: a) Erd\H{o}s-R\'{e}nyi \cite{erdos1959random}; b) Newman-Watts-Strogatz \cite{newman1999renormalization}; c) Watts-Strogatz \cite{watts1998collective}; d) Barab\'{a}si-Albert \cite{barabasi1999emergence}; e) Holme-Kim \cite{holme2002growing}.
	Nodes of the same color belong to the same community.
	Nodes are placed according with the spring layout given by the Fruchterman-Reingold algorithm \cite{fruchterman1991graph}.
	}%
	\label{fig:communities}%
\end{figure}

%*********************************************************
\section{Results of the Experimental Evaluation} \label{sec:results}
%The discussion of experimental results follows.
%The analysis of collected datasets has been conducted exploiting the functionalities of the  Stanford Network Analysis Platform library (SNAP) \cite{snap}, which provides general purpose network analysis functions.

%Data have been plotted using the open source tool Gnuplot \cite{williams1993gnuplot}.

%*******************************************
\subsection{Description of Adopted Online Social Network Datasets}
Our experimentation has been conducted on different Online Social Networks whose dataset are available online. 

Datasets $1-5$ are taken from Arxiv\footnote{Arxiv (http://arxiv.org/) is an Online archive for scientific preprints in the fields of Mathematics, Physics and Computer Science, amongst others.} datasets, as of April 2003, of papers in the field of, respectively: 1) ``Astro Physics'', 2) ``Condensed Matter Physics'', 3) ``General Relativity and Quantum Cosmology''; 4) ``High Energy Physics - Phenomenology'', and 5) ``High Energy Physics - Theory''. 
Dataset 6 represents a network of scientific citations among papers belonging to the Arxiv  ``High Energy Physics - Theory'' field.
Dataset 7 illustrates the email communications among the Federal Energy Regulatory Commission members.
Dataset 8 describes a sample of the Facebook friendship network, representing its social graph.
Dataset 9 depicts the social graph of YouTube as of 2007.
Finally, dataset 10 depicts the voting system of Wikipedia for the elections of administrators that occurred in January 2008.
Adopted datasets have been summarized in Table \ref{tab:datasets}.

\begin{table}[htp]\centering 	\small		
	\tbl{Datasets and results: $d(q)$ is the effective diameter, $\gamma$ and $\sigma$, resp., the exponent of the power law node degree and community size distributions, $Q$ the \emph{network modularity}.}	
	{\begin{tabular}{@{}c@{}c@{}c@{}c@{}c@{}c@{}c c c c c c@{}}
		\hline
		no. &	Network 		&	no. nodes \	&	no. edges	\ &	\ Dir. \ &	Type				& \ $d(q)$ \ & $\gamma$	& $\sigma$ & $Q$ &	Ref\\	
		\hline \hline
		1 & CA-AstroPh		&	18,772		&	396,160		&	No		&	Collaborat.	& 5.3			&	$2.23$ & $1.50$ & $0.628$	&	\cite{Leskovec2006}	\\
		2	&	CA-CondMat		&	23,133		&	186,932		& No		&	Collaborat.	&	7.9			&	$2.65$ & $1.49$ & $0.731$	&	\cite{Leskovec2006}	\\	
		3	&	CA-GrQc				&	5,242			&	28,980		&	No 		& Collaborat.	&	8.9			&	$2.12$ & $1.48$ & $0.861$	&	\cite{Leskovec2006}	\\	
		4	&	CA-HepPh			&	12,008		&	237,010		&	No 		& Collaborat.	&	6.6			&	$1.71$ & $1.46$ & $0.659$	&	\cite{Leskovec2006}	\\
		5	&	CA-HepTh			&	9,877			&	51,971		&	No 		& Collaborat.	&	8.4			&	$2.63$ & $1.46$ & $0.768$	&	\cite{Leskovec2006}	\\
		6	&	Cit-HepTh			&	27,770		&	352,807		&	Yes		& Citation		&	6.5			&	$3.28$ & $1.48$ & $0.658$	&	\cite{Leskovec2006}	\\	
		7 & Email-Enron		&	36,692		&	377,662		&	Yes		& Collaborat. &	5.4			&	$1.84$ & $1.48$ & $0.615$	&	\cite{Leskovec2006}	\\
		8	&	Facebook			&	63,731		&	1,545,684	&	Yes		&	Online Com.	&	6.8			&	$2.91$ & $1.48$ & $0.634$	&	\cite{mislove2007measurement}\\
		9	&	Youtube				&	1,138,499 &	4,945,382 &	Yes		&	Online Com. &	7.6			&	$2.05$ & --			& $0.447$		&	\cite{mislove2007measurement}\\
		10& Wiki-Vote			& 7,115			&	103,689		&	Yes		&	Collaborat.	&	4.5			&	$1.38$ & -- 		& $0.418$		&	\cite{Leskovec2006}	\\
		\hline
	\end{tabular}}
	\label{tab:datasets}
\end{table}

%%%%%%%%%%%%%%%%%%%%%%%%%%%%%%%%%%%
\subsection{Topological Features}
Several measures are usually needed in order to study the topological features of Social Networks.
To this purpose, for example, Carrington et al. \cite{carrington2005models} propose a list of some of them, including, amongst other, nodes/edges degree distributions, diameter, clustering coefficients, and more.

In this work, the following features have been investigated: i) node degree distribution; ii) diameter and hops; iii) \emph{community structure}.

%%%%%%%%%%%%%%%%%%%%%%%%%%%%%%%%%%%%%%%%%%%%%%%%%%%%%%%%%%%%%%%%%%%
\subsubsection{Degree distribution} \label{sub:degree-distribution}
The first interesting  feature we analyzed is the degree distribution, which reflects in the topology of the network.
The literature refers that Social Networks are usually described by power law degree distributions, $P(k)\sim k^{-\gamma}$, where $k$ is the node degree and $\gamma \le 3$.

The degree distribution can be represented by using some distribution functions: one of the most commonly used is the Complementary Cumulative Distribution Function (CCDF)

\begin{equation}
	\wp (k) = \int_k^\infty P(k')dk' \sim k^{-\alpha} \sim k^{-(\gamma -1)}.
	\label{eq:ccdf}
\end{equation}

\begin{figure*}%
	\centering
	\includegraphics[width=\columnwidth]{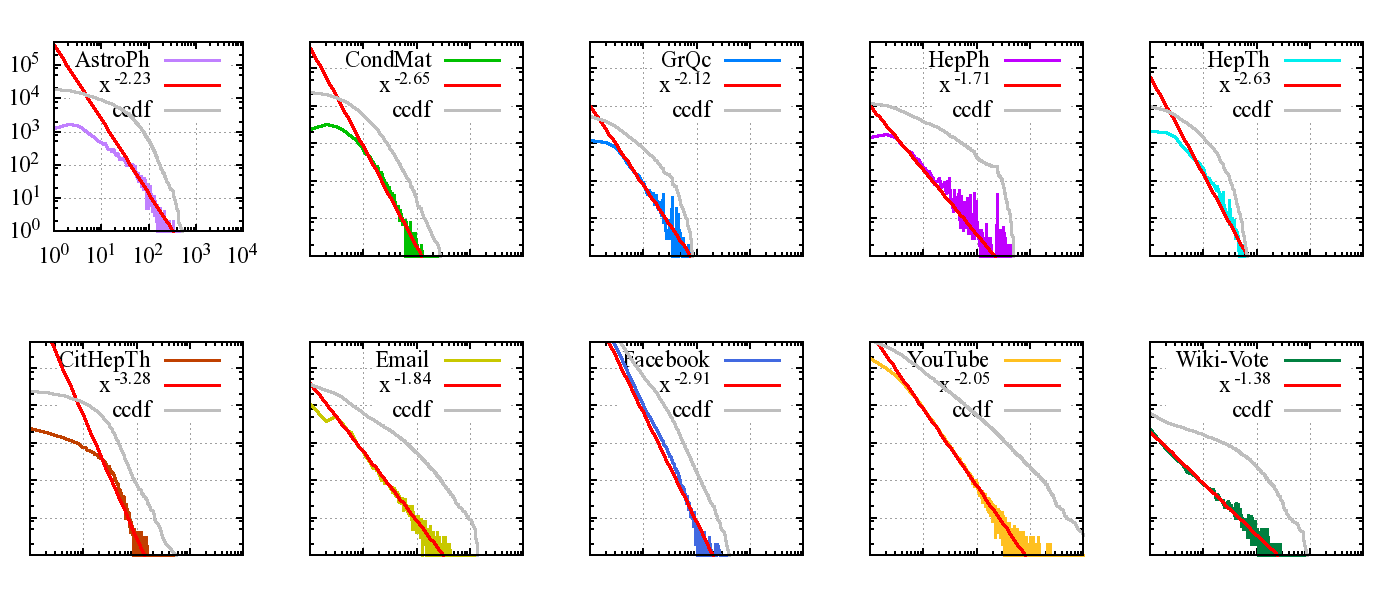}%
	\caption{Node degree distributions (\emph{log--log} scale): for each network we plot the data, the best fitting power law function and \emph{complementary cumulative distribution function} (all the plots use the same scale of the first one).}%
	\label{fig:outdegc}%
\end{figure*}

In Figure \ref{fig:outdegc} we show the degree distribution and the correspondingly CCDF evaluated on our Online Social Networks.
For those networks that are directed, the out-degree is represented.
All the graphics are represented by using a log-log scale, in order to put into evidence the scale-free behavior shown in these networks.
In particular, for each of these distributions we estimated the value of $\gamma$ in Equation \ref{eq:powerlaw}.
Values of $\gamma$ are reported in Table \ref{tab:datasets}.

Online Social Networks can be classified into two categories: i) networks that are properly described by a power law distribution; ii) networks that show some fluctuations with respect to those power law distributions that best fit to the real data.
We discuss these two categories separately.

The networks that are well described by a power law distribution, such as those depicting datasets 7--10, are all Online Communities (i.e., networks of individuals connected by social ties such as a friendship relations) characterized by a fact: the most of the users are rather inactive and, thus, they have a few connections with others members of the network.
This phenomenon shapes a very consisting and long tail and a short head of the power law distribution of the degree (as graphically depicted by the respective plots in Figure \ref{fig:outdegc}).

The former category, instead, includes the co-authors networks (datasets 1--5) that are collaboration networks and a citation network (dataset 6).
The plotting of these data against the power law distribution that try to best fit them show some fluctuation, in particular in the head of the distribution, in which, apparently, the behavior of the distribution is not properly described.
The rationale behind this phenomenon lies into the intrinsic features of these networks, that are slightly different in respect to Online Communities.

For example, regarding the co-authors networks that represent collaborations among Physics scientists, the most of the papers are characterized by the following behavior, that can be inferred from the analysis of the real data: the number of co-authors tends to increase up to 3 (in average), then it slowly slopes down to a dozen, and finally it quickly decreases.

A similar interpretation holds even for the citation network, that is usually intended as a network in which there is a very small number of papers that gather the most of the citations, and a huge number of papers that have few or even no citations at all.
This is a well-known phenomenon, called the ``first-mover effect'' \cite{newman2009first}.

Intuitively, from a modeling perspective, the only viable solution to capture these scale-free degree distribution behaviors would be by means of a Barab\'{a}si-Albert \emph{preferential attachment} model.
On the one hand, by using this model it would be possible to reproduce the power law degree distribution of all the Online Social Networks depicted above.
Similarly, even the ``Small World'' effect that describes networks with small diameters would be captured.
On the other hand, this model would fail in depicting the \emph{community structure} of those networks, whose existence has been put into evidence, both in this study (see Section \ref{sec:community-structure}) and in other works \cite{newman2001structure,girvan2002community,leskovec2008statistical}.

%%%%%%%%%%%%%%%%%%%%%%%%%%%%%%%%%
\subsubsection{Diameter and hops}
Most real-world Social Networks exhibit a relatively small diameter, but the diameter is susceptible to outliers.
A more reliable measure of the pairwise distances between nodes in a graph is the \emph{effective diameter}, i.e., the minimum number of hops in which some fraction (or quantile q, say $q = 0.9$) of all connected pairs of nodes can reach each other.
The \emph{effective diameter} has been found to be small for large real-world networks, like Internet and the Web \cite{albert1999diameter}, real and Online Social Networks \cite{albert2002statistical}.

A hop-plot extends the notion of diameter by plotting the number of reachable pairs $g(h)$ within $h$ hops, as a function of the number of hops \emph{h}.
It gives us a sense of how quickly neighborhoods of nodes expand with the number of hops.

\begin{figure}%
	\centering
	\includegraphics[clip=true, trim = 0mm 35mm 0mm 35mm, width=.75\columnwidth]{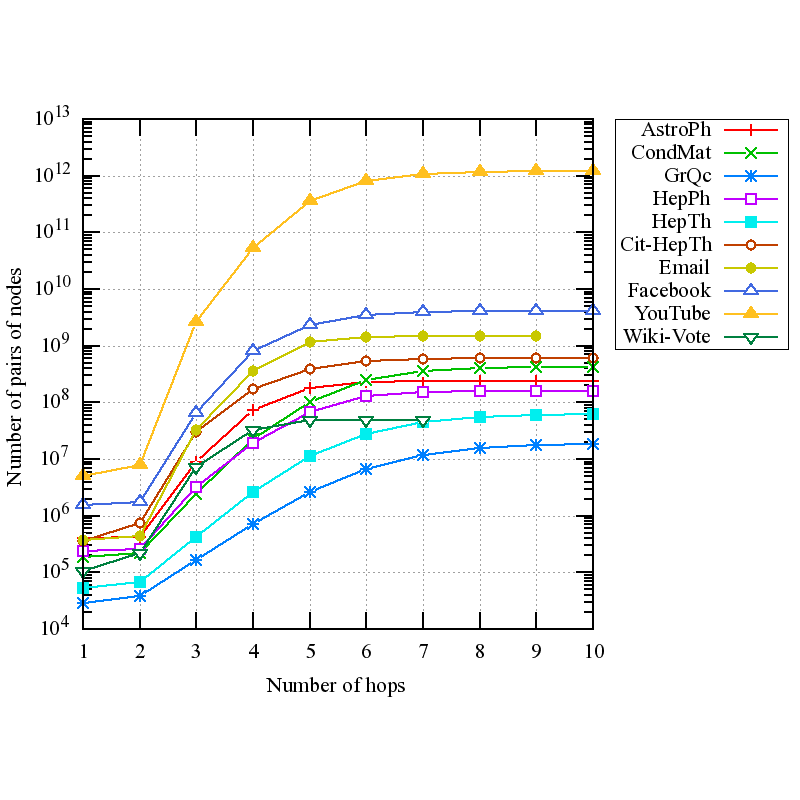}%
	\caption{Effective diameters (\emph{log-normal} scale): the number of pairs of nodes reachable is plotted against the number of hops required, with $q=0.9$.}%
	\label{fig:hop}%
\end{figure}

In Figure \ref{fig:hop} the number of hops necessary to connect any pair of nodes is plotted as a function of the number of pairs of nodes, for each given network.
As a consequence of the \emph{compact structure} of these networks (highlighted by the scale-free distributions and the ``Small World'' effect, discussed above), diameters show a fast convergence to asymptotic values listed in Table \ref{tab:datasets}.

From a modeling standpoint, as for the degree distributions, the previous considerations hold true.
Both the Watts-Strogatz and the Barab\'{a}si-Albert models could efficiently depict the ``Small World'' feature of these Online Social Networks, and, most importantly, empirical data verify the so called ``Six degrees of separation'' theory, that is strictly related with the ``Small World'' formulation.
In fact, it is evident that, regardless of the large scale of the networks analyzed, the effective diameters are really small (and, in average, close to 6), which is proved for real-world Social Networks \cite{watts1998collective,barrat2000properties}.

%*******************************
\subsubsection{Community Structure} \label{sec:community-structure}
From the perspective of the models representing the \emph{community structure} of a network, we can infer the following information: from Figure \ref{fig:communities}.a, where the \emph{community structure} of a Erd\H{o}s-R\'{e}nyi model is represented, the result appears random, according to the formulation of the model and its expected behavior when the calculated \emph{network modularity} $Q$ function (Equation \ref{eq:qmod}) is analyzed.
From Figures \ref{fig:communities}.b-c, at a first glance, it emerges that the \emph{community structure} of Watts-Strogatz models is very regular and there is a balance between communities with tighter connections and those with weaker connections. 
This reflects the formulation of the model but does not well depict the \emph{community structure} represented by scale-free networks.
Finally, Figures \ref{fig:communities}.d-e appear more compact and densely connected, features that are not reflected by experimental data.

Even if well-representing the ``Small World'' effect and the power law distribution of degrees, the Barab{\'a}si-Albert model and its variants appear inefficient to represent the \emph{community structure} of Online Social Networks.

From our experimental analysis on real datasets, by analyzing the obtained \emph{community structures} by using the \emph{Louvain method} \cite{blondel2008fast}, we focus on the study of the distribution of the dimensions of the obtained communities (i.e., the number of members constituting each detected community) \cite{fortunato2010community}.

Recently Fortunato and Barthelemy \cite{fortunato2007resolution}, put into evidence a resolution limit while adopting \emph{network modularity} as maximization function for the community detection.
In detail, authors found that modularity optimization may fail in the discovery of communities whose size is smaller than a given threshold.
This value is strictly correlated to the characteristics of the given network.
This resolution limit results in the creation of large communities incorporating an important part of the nodes of the network.
In practice, in some particular cases it is possible that the clustering process produces a small number of communities of big sizes.
This would possibly affect results in two ways: i) enlarging the tail of the power law distribution of the size of the community, or ii) producing a not significant clustering of the network.

Because the clustering algorithm adopted (i.e., the \emph{Louvain method}) is a modularity maximization technique, we investigated the effect of the resolution limit on our datasets.
We found that in two cases (i.e., for the datasets 9--10) the clustering obtained was biased from the resolution limit and we excluded these networks from our analysis.

In the following we investigate the behavior of the distribution of the size of the communities in our networks.

\begin{figure}%
	\centering
	\includegraphics[width=\columnwidth]{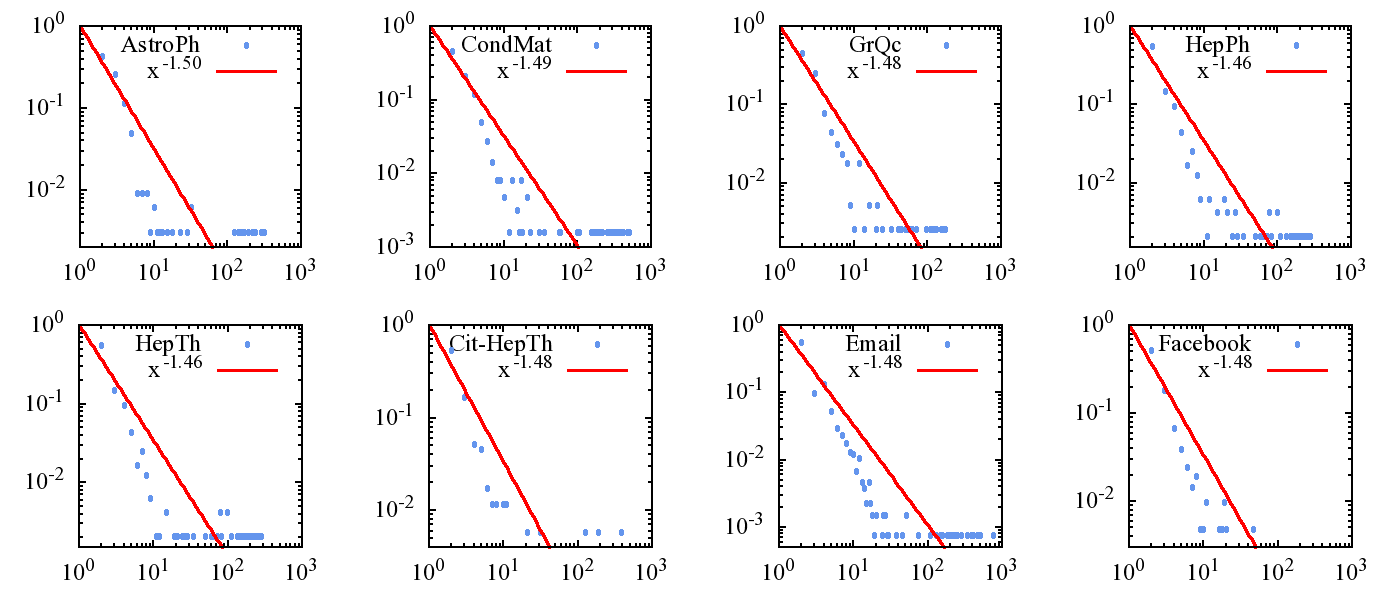}%
	\caption{\emph{Community structure} analysis (\emph{log--log} scale): the probability $P(x)$ of finding a community of a given size into the network is plotted against the size of the community. In red, the best fitting power law distribution functions are depicted.}%

	\label{fig:community-size}%
\end{figure}

In Figure \ref{fig:community-size} on the x-axis we plot the size of the community, and on the y-axis the probability P(x) of finding a community of the given size into the network.
For each distribution we provide the best fitting power law function with a given exponent $\sigma$ (that always ranges into the interval [1.4,1.5]) that well approximates the behavior of the community size.

In the figure the data are plotted as points and it is possible to highlight some communities whose size is larger than that expressed by the expected power law function (plotted as a red line), that constitute the heavy tail of the power law distribution.
The results depicted show that, inside large Online Social Networks, there is a high probability of finding a high number of communities that contain few individuals and a relatively low probability of finding communities constituted by a large number of members.
This confirms that individuals are more likely to aggregate in small communities, such as those representing family, friends, colleagues, etc., rather than in large communities \cite{porter2009communities,ijsnm2011}.

Moreover, from Figure \ref{fig:community-size} we can put into evidence that large Online Communities, for example Facebook and the scientific collaboration networks, show a very tight \emph{community structure} (a fact proved also by the high values of \emph{network modularity}, reported as $Q$ in Table \ref{tab:datasets}).
For example, regarding the collaboration networks, intuitively, we interpret this fact considering that scientists usually co-authoring different works, with different persons, work on papers signed only by a small amount of co-authors.
It is very likely that these co-authors tend to group together (for example, if they co-authored several works) in the corresponding scientific communities.

On the other hand, for some networks such as the citation network and the email network, Figure \ref{fig:community-size} shows that it exists an important amount of communities constituted by a large amount of individuals, constituting the heavy long tail of the power law distribution.
Also this aspect has an intuitive explanation.
In fact, if we consider a network of scientific citations, there is a small amount of papers with a huge number of citations (which are very central in the topology of the network and, thus, are aggregated in same communities) and the most of the others that have very few citations, that forming small communities among each other (or single entities).

%********************************************
\section{Conclusions} \label{sec:conclusions}
In this paper we put into evidence those models which try to efficiently and faithfully represent the topological features of a family of complex networks, called Online Social Networks. 

Several models have been presented in literature and we focused our attention on the three main exploited models, i.e., i) Erd\H{o}s-R\'{e}nyi random graphs, ii) Watt-Strogatz and, iii) Barab{\'a}si-Albert preferential attachment.
Each model, even if well describes some specific characteristics, fails in faithfully representing all the three main features we identified, that characterize Online Social Networks, namely i) ``Small World'' effect, ii) \emph{scale-free degree distributions} and, finally, iii) emergence of a \emph{community structure}.

We analyzed the topological features of several real-world Online Social Networks, fitting real data to models and putting into evidence what characteristic are preserved and what could not faithfully be represented by using these models.

As for future work, our main aim is to provide a generative model that, at the best of our knowledge, for first would try to faithfully represent the three principal features of Online Social Networks we have identified.

%**************************
\section*{Acknowledgements}
We would like to thank the Editor and the anonymous Referees whose comments helped us to greatly improve the quality of the work.

\bibliographystyle{CAIMbibstyle}
\bibliography{CAIM-bib}

\end{document}